\documentclass[usenatbib]{basi}
\usepackage[british]{babel}
\usepackage[varg]{txfonts}
\usepackage{rotating}
\usepackage{dcolumn}
\begin{document}
\title[The gamma-ray bursts, core-collapse supernovae 
and  global star forming rate at large redshifts]{The gamma-ray bursts, core-collapse supernovae 
and  global star forming rate at large redshifts }
\author[ V.~V.~Sokolov]%
       { V.~V.~Sokolov$^1$\thanks{email: \texttt{sokolov@sao.ru}}\\ 
      $^1$ Special Astrophysical Observatory of the Russian Academy 
      of Sciences (SAO RAS), Nizhnij Arkhyz 369167, Russia\\
       }

\pubyear{2011}
\volume{00}
\pagerange{\pageref{firstpage}--\pageref{lastpage}}

\date{Received \today}

\maketitle
%------------------------------------------------------------------------------%
% abstract and keywords                                                        %
%------------------------------------------------------------------------------%
\label{firstpage}

\begin{abstract}
The brief review and discussion of statement of some observational problems of gamma-ray bursts (GRB), GRB host galaxies and star forming at small and large redshifts: Are there similarities and differences between GRB hosts and the typical galaxy population -- this is currently the main question for the study of GRB host galaxies. The direct connection between long-duration GRBs and massive stars explosions, GRBs and some puzzles of core-collapse supernovae are briefly discussed. On model-independent observational cosmological tests -- GRB rate and star forming rate at high redshifts.
\end{abstract}

\begin{keywords}
  gamma-ray bursts: star formation – galaxies: supernovae 
\end{keywords}

%------------------------------------------------------------------------------%
% main text of the paper, using \section, \subsection, \subsubsection          %
%------------------------------------------------------------------------------%

% ----------- Section 1. Introduction ----------------------
\section{Introduction}\label{s:intro}

Since the first optical identification of gamma-ray bursts (GRBs) in 1997 \citep{Costa1997, Paradijs1997}, they became a new direction in the study of the universe at large redshifts. In particular, on April, 23, 2009 the redshift $z=8.2$ was measured, and this happened to be of GRB 090423 (\citep{Tanvir2009, Salvaterra2009}, see also \citep{Cucchiara2011} on the photometric $z \sim 9.4$ for GRB 090429B).  
From 1997 to 2011, the general state of the GRB problem and progress in this field can be fixed in the following way: 1) GRBs belong to the most distant observable objects with measurable redshifts in the universe, 2) GRBs are related to the starforming in distant (and very distant) galaxies. 3) GRBs and their afterglows allow us seeing also the most distant explosions of massive stars at the end of their evolution,  4) This is confirmed by observations for long-duration bursts, but most probably short GRBs are also related to some very old compact objects resulting from evolution of the same massive stars.

The aim of the paper is the short review and discussion of some urgent observational problems related to GRBs. The outline of the paper is as follows: Section 2 concerns the optical identification -- the first GRB host galaxies and (massive) star forming rate (SFR) at smaller redshifts, the metallisities of GRB hosts and the question if there are similarities and differences between GRB hosts and normal field galaxies at larger redshifts. The direct connection between long-duration GRBs and massive stars explosions, GRBs and some puzzles of core-collapse supernovae (CCSNe) are presented in Section 3. Section 4 deals with the discussion of SFR and GRB rate (GRBR) at high redshifts: is there a fast decrease in SFR up to $z \sim 10$? This section ends in some conclusions about the study of GRB host galaxies, GRBs and CCSNe related to them and about new possible crutial cosmological tests  (on the GRB-CCSNe tests) at high redshifts. 

%Conclusions: SFR $\sim$ GRBR $\sim$ CCSN rate, GRBs and CCSNe at $z \sim 10$? 

% ---------- Section 2 -----------------------------------
\section{The optical identification:  the first GRB host galaxies  and  massive SFR}

The first X-ray and optical afterglows were observed for the very first time by \citep{Costa1997} and  \citep{Paradijs1997} for GRB 970228 with the Italian-Dutch satellite BeppoSAX \citep{Boella1997},  thank to the fast and accurate positioning of GRB obtainable through the combined capabilities of the GRB Monitor and Wide Field Cameras onboard this famous satellite. Optical observations of the following GRB 970508 optical remnant were continued with the 6-m telescope of SAO RAS in standard {\it $BVR_cI_c$} bands in Oct.-Dec. 1997 and in Jan. 1998. The results of the {\it $BVR_cI_c$}  photometry for GRB 970508 optical afterglow and for three nearby galaxies was presented by Zharikov et al. (1998). Further multi-wavelength observations of GRBs have confirmed that a significant fraction of long-duration GRBs are associated with the collapse of short-lived massive ($\sim 30 M_{\odot}$) stars \citep{Hogg1999, Bloom2001}. The regions of the massive star-forming are seen in rest frame UV part spectra of star-forming galaxies. As it has turned out, it is just a light of the massive stars in the GRB hosts  \citep{Sokolov1999}. 

\begin{figure}
\centerline{\includegraphics[width=9cm]{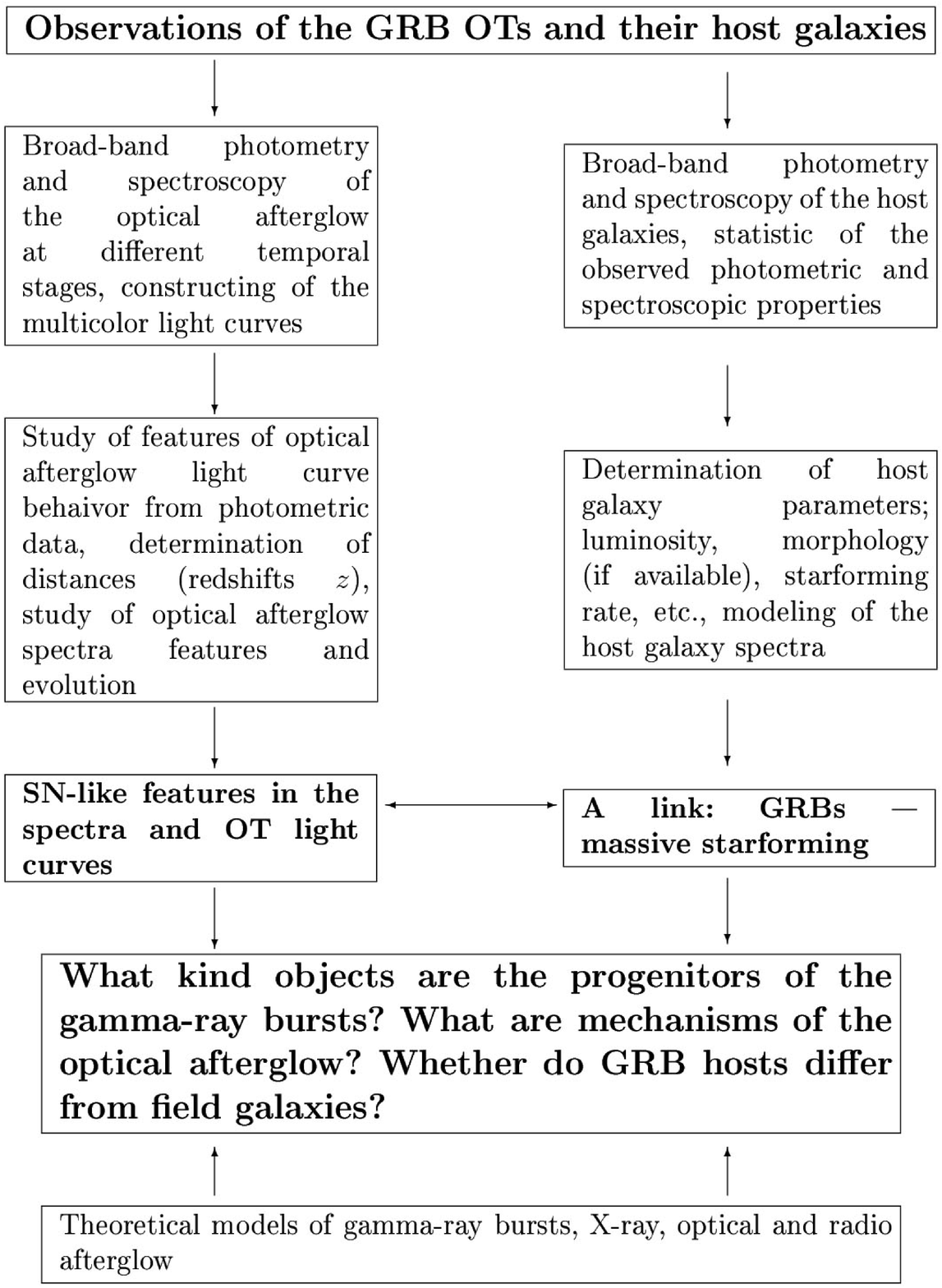}}
\caption{The astronomy of $\gamma$-ray bursts  with the 6-m telescope from 1998
\label{f:Fig1}}
\end{figure}

Fast localization with BeppoSAX and ground based follow-up optical observations with the  measurements of GRB redshifts have shown their (GRBs) relation to distant galaxies located in sites of faded transients. This was essentially {\it the first stage of optical identification} with massive star-forming galaxies -- the objects with more or less clear properties in contrast to properties of the GRB optical afterglow emission which are obscure thus far. It is the study of these objects (the broad-band photometry and spectroscopy of the GRB hosts, statistic of the observed photometric and spectroscopic properties and so on) that launched the GRB astronomy with the 6-m telescope in 1998. At that time researchers studying this side (see Figure \ref{f:Fig1}) of GRB optical counterparts were interested mostly in the question {\it whether GRB hosts differ from field galaxies}. Then there were ideas that such unique sources as the GRBs could be related for sure to some unusual (unique) objects of type of quasars, or something like that.   

%---- Figure 1 ---------------------
 
\begin{figure}
\centerline{\includegraphics[width=9cm]{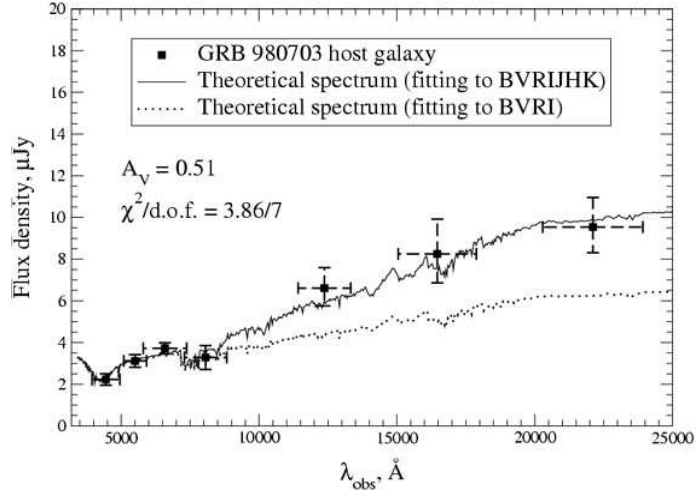}}
\caption{The population synthesis modelling: Comparison of modelled and observed fluxes in the filters {\it $B, V, R_c, I, J, H, K$} for the GRB 980703 host galaxy ($z=0.9662$). If GRBs are associated with an active star formation, then we might expect the light of their host galaxies to be affected by {\it internal extinction} \citep{Sokolov2001a}.
\label{f:Fig2}}
\end{figure}

The study of physical properties of host galaxies \citep{Sokolov1999} permits determining the differences from usual galaxies (in the same CCD fields like for GRB 070508) with massive star-forming, which gives us a key to the understanding of conditions in which a GRB progenitor object is born, evolves and dies. But the most distant host galaxies can be often observed only photometrically. In these cases, such physical properties as SFR, intrinsic extinction laws, ages, masses and metallicities can be estimated only by a population synthesis modelling \citep{Sokolov2001a} of spectral energy distributions (SEDs). In the case of the host galaxy of GRB 980703 ($z=0.9662$) the observed deficit in the B-band (see Figure \ref{f:Fig2}) can be explained by the excess of extinction near 2200\AA{}, which is characteristic of the extinction law similar to that of the Milky Way. 
And the recent confirmation \citep{Zafar2011} of the extinction near 2200\AA{}  for GRB hosts with {\it the larger redshifts:} e.g. the GRB 070802 $(z = 2.4541)$ afterglow which was observed with the VLT/FORS2 and the optical spectrum  of GRB 080607 afterglow ($z =3.0368$) which was obtained with the Keck telescope. The 2175\AA{} dust extinction feature is clearly seen in the optical spectra of the afterglows (see e.g.  Fig.A.2 in \citep{Zafar2011}).

%---- Figure 2 ---------------------

%--------- Table ! ---------------
\newcolumntype{d}[1]{D{.}{.}{#1}}
\begin{table}
  \caption{The selected parameters of two host galaxies.}\label{tab:simple}
  \medskip
  \begin{center}
  \begin{small} %---my --
    \begin{tabular}{cccccccc}\hline \hline 
        Host             & Scenario            & Metallicity           & Total mass                  & Age      & $A_V$ & Observed SFR$^\mathrm{*}$ & Corrected SFR \\\hline
        GRB 970508 & instant.burst     & $0.1Z_{\odot}$ & $3.48\cdot10^8$      & 160Myr & 1.6      & $\ge1.4M_{\odot}yr^{-1}$ & $14M_{\odot}yr^{-1}$   \\
        GRB 980703 & exp.decreasing & $Z_{\odot}$      & $3.72\cdot10^{10}$ & 6 Gyr    & 0.64    & $\ge10M_{\odot}yr^{-1}$ & $20M_{\odot}yr^{-1}$  \\\hline
    \end{tabular}\\[5pt]
    \begin{minipage}{14cm}
      \small $^\mathrm{*}$ The SFR was recomputed following cosmology with $H_0=60 km\cdot s^{-1}\cdot Mpc^{-1}$, $\Omega_M=0.3$ and $\Omega_{\Lambda}=0.7$ \\
    \end{minipage}
   \end{small} %--- my --
  \end{center}
\end{table}
% --------------------------------------------

So, in two cases (see Table \ref{tab:simple}) of pure host galaxy spectra: GRB 970508 ($M_{B_{rest}} = -18.62$) and GRB 980703 ($M_{B_{rest}} = -21.27$) we performed the theoretical modeling of the continuum spectral energy distribution of these hosts, making use of the spectra and our $BVR_cI_c$ photometry \citep{Sokolov2001b}. These two examples demonstrate that $BVR_cI_c$ photometrical spectral distribution describes well the spectral continua of the host galaxies. Furthermore, in the case of absence of the spectra of a host galaxy itself, the broad-band photometry is the only tool to study spectral energy distribution for these very distant faint galaxies with the integral $25-26^{th}$ magnitude and fainter (see Figures \ref{f:Fig4} and \ref{f:Fig5}). 
       In fact, the identical method of the broad band SEDs fitting (and determination of redshift, luminosity, stellar mass, age, metallicity and other parameters) is widely used in recent studies of very distant and faint (28-th magnitude) galaxies involving results of the broad-band photometry in near and middle IR. The example of best-fit stellar population models at $z \sim 7$ are shown in the paper: ''The evolution of $z=7-8$ galaxies from IRAC observations of the DEEP/WIDE-AREA WFC3/IR ERS and ULTRADEEP WFC3/IR HUDF'' (see Fig.1 in \citep{Labbe2009}). 

%---- Figure 3 ---------------------
\begin{figure}
\centerline{\includegraphics[width=11cm]{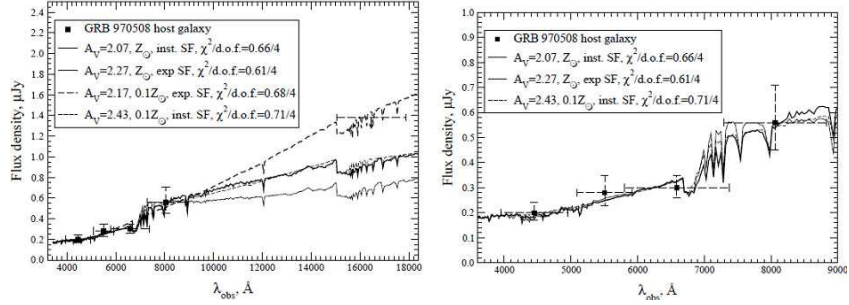}}
\caption{The best fit for the spectral energy distribution (SED) model to the $BVR_cI_c$ photometry of the GRB 970508 host galaxy $(z=0.8349)$, assuming the Calzetti extinction law. Also the upper limit of HST/NICMOS $H$-band is plotted (Left). The observed wavelengths are given in \citep{Sokolov2001a}. 
\label{f:Fig3}}
\end{figure}

%---- Figure 4 ---------------------
\begin{figure}
\centerline{\includegraphics[width=9cm]{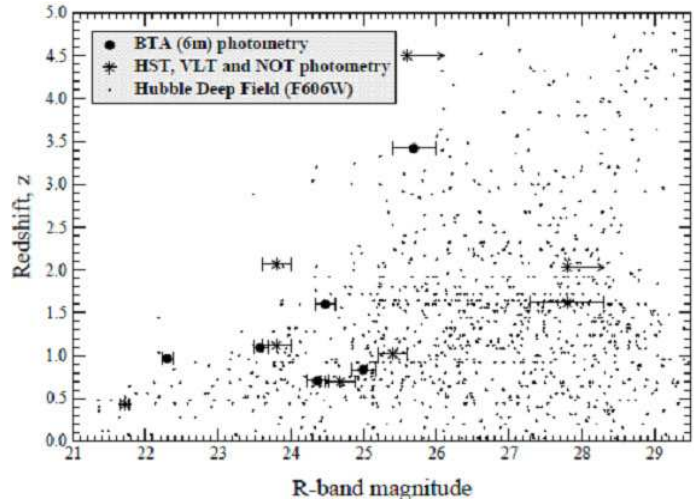}}
\caption{The observed $R$-band magnitude vs. spectroscopic redshift for the first 12 GRB host galaxies.  The BTA $R$-band magnitudes (from \citep{Sokolov2001a}) are marked with circles, while asterisks refer to the results of other authors. Also the HDF F606W magnitude vs. {\it photometrical redshift} distribution is plotted. Catalog of the F606W magnitudes and photometrical redshifts was used from Fern\'andez et al.(1999).
\label{f:Fig4}}
\end{figure}

%---- Figure 5 ---------------------

\begin{figure}
\begin{center}
\begin{tabular}{p{6cm}cp{6cm}}
\raisebox{-\height}{\includegraphics[width=5.5cm]{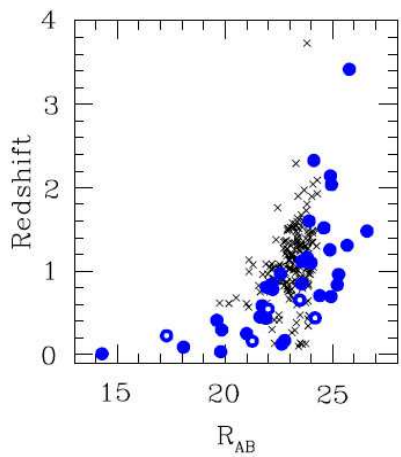}} & \quad &
\caption{The observed magnitudes as a function of redshift from the paper by Savaglio et al.(2008)  for GRB hosts (filled circles) and  Gemini Deep Survey field galaxies (crosses,Abraham et al.(2004)). The filled circles with white dots are short-GRB hosts.\label{f:Fig5}}
\end{tabular}
\end{center}
\end{figure}

        To summarize the results of the GRB host modeling \citep{Sokolov2001a, Sokolov2001b}, we can conclude the following. (i) Broad-band flux spectra of GRB host galaxies are well fitted by SEDs of {\it the local starburst galaxies}. (ii) UV part of GRB host galaxy SEDs is properly described by theoretical models with {\it young burst} star formation. Moreover, for $z \sim 1$ in the optical waveband, in fact, we observe only star-forming regions in GRB host galaxies, because they dominate the rest-frame UV part of the spectrum. (iii) It is important to take into account the effects of the {\it internal extinction} in the modeling of GRB host galaxy SEDs. (iv) GRB host galaxies seem to be the same as galaxies at the same redshift. 

         So, we have concluded that long-duration GRBs seem to be closely related to vigorous massive star-forming in their host galaxies. It should be noted that the SFR in the host galaxies is unlikely to be much higher than in galaxies at the same redshifts ($z \gtrsim 1$). At  this redshift the mean star formation rate is $\sim20-60 M_{\odot} yr^{-1}$ (see also Blain et al.(2000)). For these reasons we conclude that GRB host galaxies seem to be similar to field galaxies at the same redshift \citep{Sokolov2001a, Sokolov2001b}. And at present we have the independent confirmation by Savaglio et al.(2006): GRBs are identified with ordinary galaxies indeed. In the paper was used of GRB Host Studies public archive collecting observed quantities of 32 GRB host galaxies., i.e. about half of the total number of GRBs with redshift known by January 2006. The authors present some preliminary statistical analysis of the sample, e.g. the total stellar mass, metallicity and SFR for the hosts. The total stellar mass and the metallicity for a subsample of 7 hosts at $0.4<z<1$ are consistent with the mass-metallicity relation found for normal star-forming galaxies in the same redshift interval. 

And in a later study (\citep{Savaglio2008} and references therein) formulated more definite conclusions but for the {\it larger redshift} GRB hosts: There is no clear indication that GRB host galaxies belong to a special population. Their properties are those expected for normal star-forming galaxies, from the local to the most distant universe. Metallicities measured from UV absorption lines in the cold medium of GRB hosts at $z>2$ (GRB-DLAs) are in a similar range. Combining this with the results for $z<1$ GRB hosts, we see no significant evolution of metallicity in GRB hosts in the interval $0<z<6$. And the outcome in the paper by Savaglio et al.(2008): GRB hosts should not be special, but normal, {\it faint}, star-forming galaxies (the most abundant), detected at any $z$ just because a GRB event has occurred. 

And again on metallisities and the GRB hosts \citep{Mannucci2011}: Many recent studies have attempted to find similarities and diferences between the GRB host population and the normal field galaxy one (see, for example, Fynbo et al.(2008). In particular, these studies compared the observed mass-metallicity relation (or luminosity-metallicity relation) of the two populations.  From the analysis of a whole sample of known GRB hosts, Savaglio et al.(2009) concluded again that there is no clear indication that GRB host galaxies belong to a special population. Their properties are those expected for normal star-forming galaxies, from the local to the most distant universe. Mannucci et al.(2011) have compared the metallicity properties of a sample of 18 GRB host galaxies with those of the local field population.  In particular, they have found that GRB hosts do follow the Fundamental Metallicity Relation (FMR) recently found by \citep{Mannucci2010a}. {\it This fact implies that GRB hosts do not differ substantially from the typical galaxy population.} The typical low, sub-solar metallicity found in many recent studies (e.g., Savaglio et al. 2009, Levesque et al. 2010), and references therein) does not necessary mean that GRBs occur in special, low metallicity galaxies, and that a direct link between low metallicity and GRB production exists. The low metallicities of observed long GRB hosts is a consequence of the high star formation environment. The SFR appears to be the primary parameter to generate GRB events \citep{Campisi2011}.

%------------- Section 3 -------------------------
\section{The direct connection between long-duration  GRBs and massive stars (GRB--CCSN)}

Thus, from the aforesaid, it follows that there are multiple long lines of evidence that long-duration ($\sim 1s-100s$) GRBs are associated with collapse of massive stars, occurring in regions of active star formation embedded in dense clouds of dust and gas. But at present it can be already said with confidence about a direct relation between GRBs and massive progenitor stars of CCSNe. Identification of GRB 980425 with CCSN SN1998bw \citep{Galama1998} is considered to be the first one. Correspondingly, since 1998 another direction (parallel to the study of GRB hosts) was observation of mysterious optical transients (OTs) related with GRBs \citep{Sokolov1998, Zharikov1999}. From the very beginning the main aim of these observations was  (see Figure \ref{f:Fig1}) the study of the photometric features in GRB optical afterglow light curves connected with an underlying SN \citep{Sokolov2001c}. 

On the one hand, as was said above, from the observational site, the GRB/CCSN picture was indirectly supported already (e.g., Frail et al. 2002, Sokolov et al. 2001a) by the fact that all GRB hosts are star-forming, and in some cases even star-bursting galaxies. And on the other hand indeed, some GBRs have shown rebrightening and flattening in their late optical afterglows, which have been interpreted as emergence of the underlying SN light curve \citep{Zeh2004}. The key finding is photometric evidence of a late-time bump in all afterglows with a redshift $z \lesssim 0.7$ (including GRBs 030329 and 031203 with the spectroscopic confirmed GRB/CCSN connection, see below). 

For larger redshifts the data is usually not of sufficient quality, or the SN is simply too faint, in order to search for such a feature in the late-time afterglow light curve. (In addition, the rebrightening in late optical afterglows for the large $z$ is to be observable already in near IR.) This extra light is modeled well by a SN component, peaking $(1+z)(15...20)$ days after a burst. This, together with the spectral confirmation of SN light in the afterglows of GRB 021211, 030329, and 031203 further supports the view that in fact all long-duration GRBs show SN bumps in their late-time optical afterglows. Given the fact that a strong late-time bump was also found for (X-ray flash) XRF 030723 \citep{Fynbo2004} and for XRF 020903 (with spectroscopic confirmation of underlying SN light \citep{Soderberg2005} might indicate that this conclusion holds also for XRFs. So, a systematic study on the GRB afterglows with this approach suggests that all long-duration GRBs are associated with SNe \citep{Zeh2004}.
Now the GRBs and SNe with spectroscopically confirmed connection are known: GRB 980425/SN 1998bw ($z=0.0085$), GRB 030329/SN 2003dh ($z=0.1687$), GRB 031203/SN 2003lw (z=0.1055), GRB/XRF 060218/SN2006aj ($z=0.0335$), GRB 100316D/SN2010bh ($z=0.059$)  (see also the review by Hjorth et al.(2011) and references therein). Besides, it is possible to include to this list such objects as XRF 080109/SN2008D ($z=0.0065$, see below). Figures \ref{f:Fig6}-\ref{f:Fig9} show results of spectroscopic observations of some objects from the list with the 6-m telescope.

%---- Figure 6 ---------------------
\begin{figure}
\centerline{\includegraphics[width=10cm]{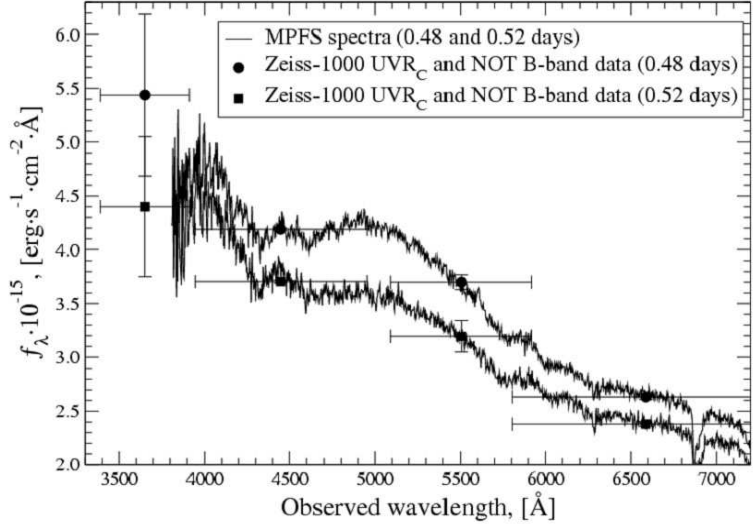}}
\caption{The earliest spectroscopy of the GRB 030329 afterglow with the 6-m telescope (MPFS/BTA), and the photometry with Zeiss-1000 \& NOT (from \citep{Sokolov2003, Kurt2005}). 
\label{f:Fig6}}
\end{figure}

%---- Figure 7 ---------------------
\begin{figure}
\centerline{\includegraphics[width=10cm]{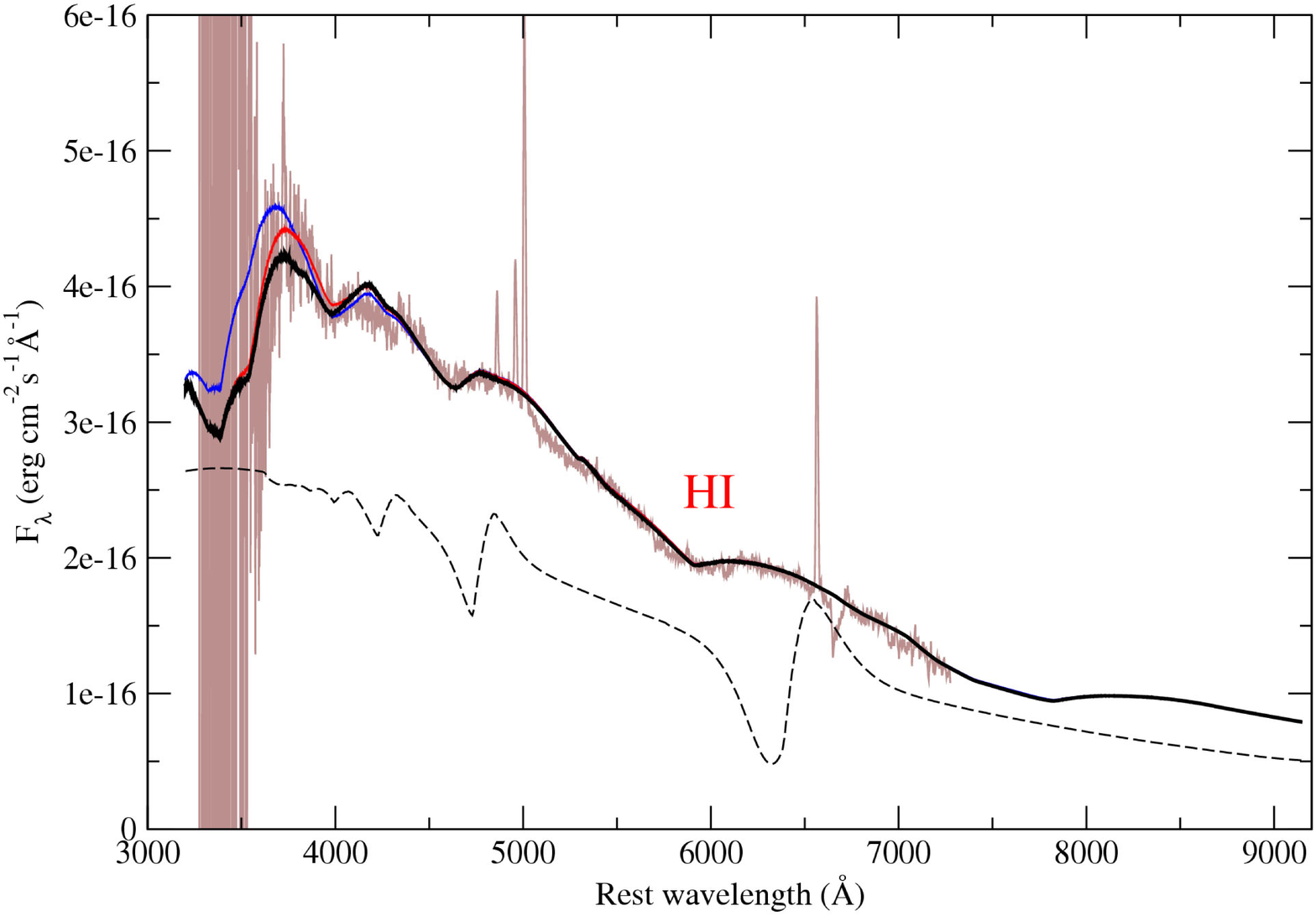}}
\caption{The SN 2006aj spectrum in rest wavelengths obtained with BTA in 2.55 days after XRF/GRB 060218 \citep{Sonbas2008}. The fitting by synthetic (SYNOW: \citep{Branch2001, Elmhamdi2006}) spectra with the velocity of the photosphere ($V_{phot}$), all elements and their ions equal to $33,000 km s^{-1}$ is shown by smooth lines differing only in the blue range of the spectrum at $\lambda < 4000$\AA{}.  HI denotes the $H_{\alpha}$ PCyg profile at $V_{phot} = 33,000 km s^{-1}$. A model spectrum for the photosphere velocity $8000 km s^{-1}$ is shown for example by the dashed line as an example of the $H_{\alpha}$ PCyg profile.
\label{f:Fig7}}
\end{figure}

%---- Figure 8 ---------------------
\begin{figure}
\centerline{\includegraphics[width=10cm]{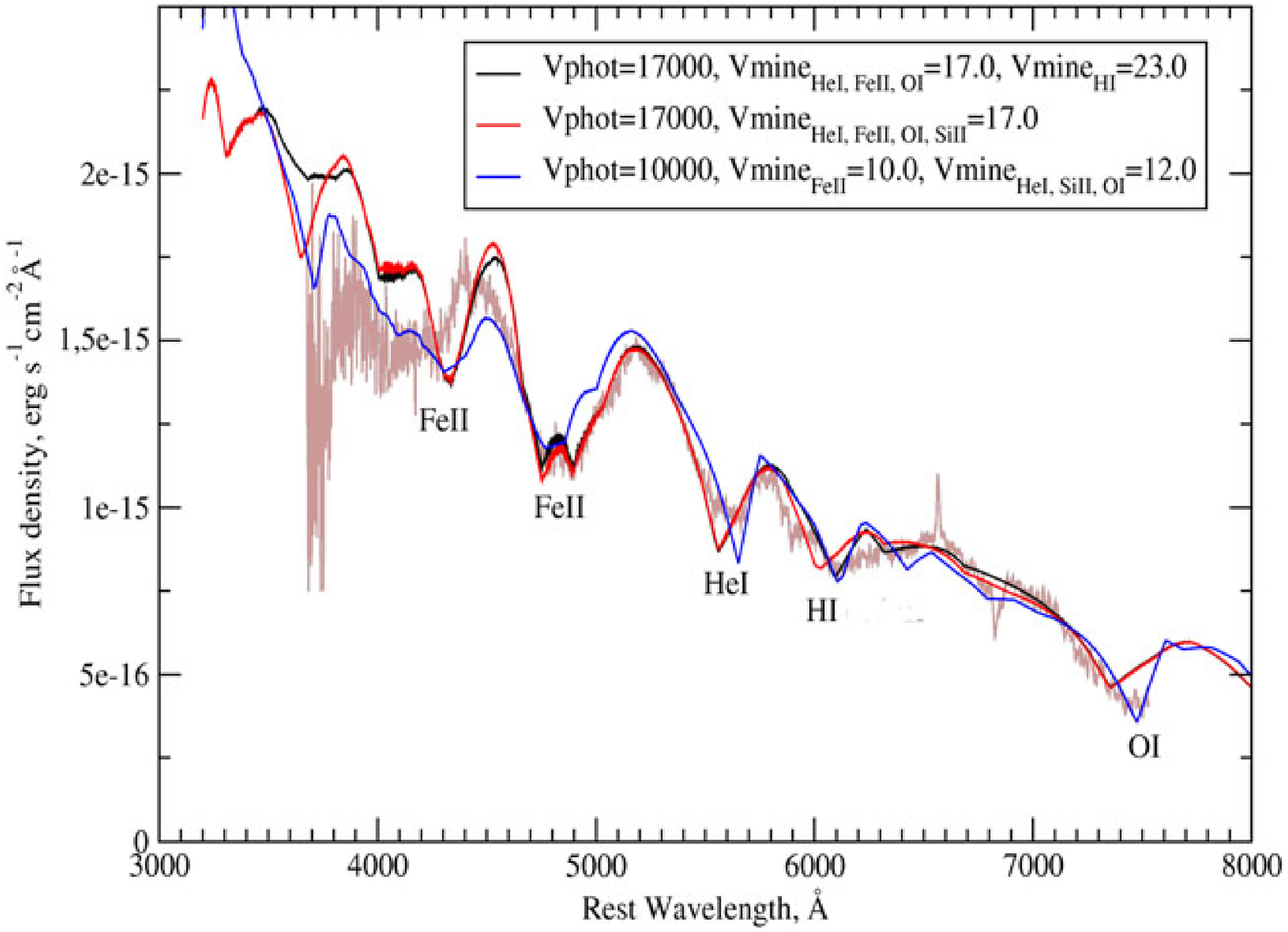}}
\caption{The spectrum of XRF080109/SN 2008D (from \citep{Moskvitin2010}). Physical conditions in the envelope of this SN were modelled with the SYNOW code \citep{Branch2002}.
\label{f:Fig8}}
\end{figure}

Thereby, long durations GRBs may be the beginning of CCSN explosion, and GRB is a signal allowing us to catch the SN at the very beginning of the explosion. At least it seams that the closer GRB has the more features of SN – nearby GRBs can show spectroscopic signs of SNe.  And though the phenomenon (GRB) is unusual, but the object-source (SN) is not too unique (this is similar to the situation with GRB host galaxies, see before). 
A popular conception of the relation between long-duration GRBs and CCSNe is shown in Figure 10. A narrow gamma-ray emission (GRB) is observed along the SN explosion axis, but closer to the equatorial plane we can observe mainly only an isotropic XRF related to the shock break out effect. This can explain why most local SNe do not show any GRB, though they show a powerful and short XRF (as was in the case of XRF 080109/SN2008D). The probability of getting into a narrow beam of gamma-ray quanta decreases as the Lorentz factor $\Gamma$ increases (the picture from Woosley et al.(2006)).

%---- Figure 9 ---------------------
\begin{figure}
\centerline{\includegraphics[width=9cm]{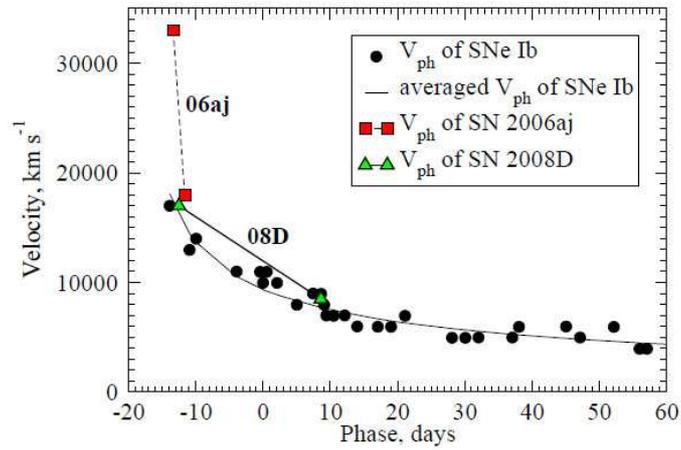}}
\caption{Velocity at the photosphere, as inferred from Fe II lines, is plotted against time after maximum light. The line is a power-law fit to the data, with SN 1998dt at 32 days (open circle) excluded (see Figure 22 in Branch et al.(2002)). Squares (SN 2008D) and diamonds (SN 2006aj) are photospheric velocities, inferred from our spectra \citep{Moskvitin2010}.
\label{f:Fig9}}
\end{figure}

%---- Figure 10 ---------------------
\begin{figure}
\centerline{\includegraphics[width=9cm]{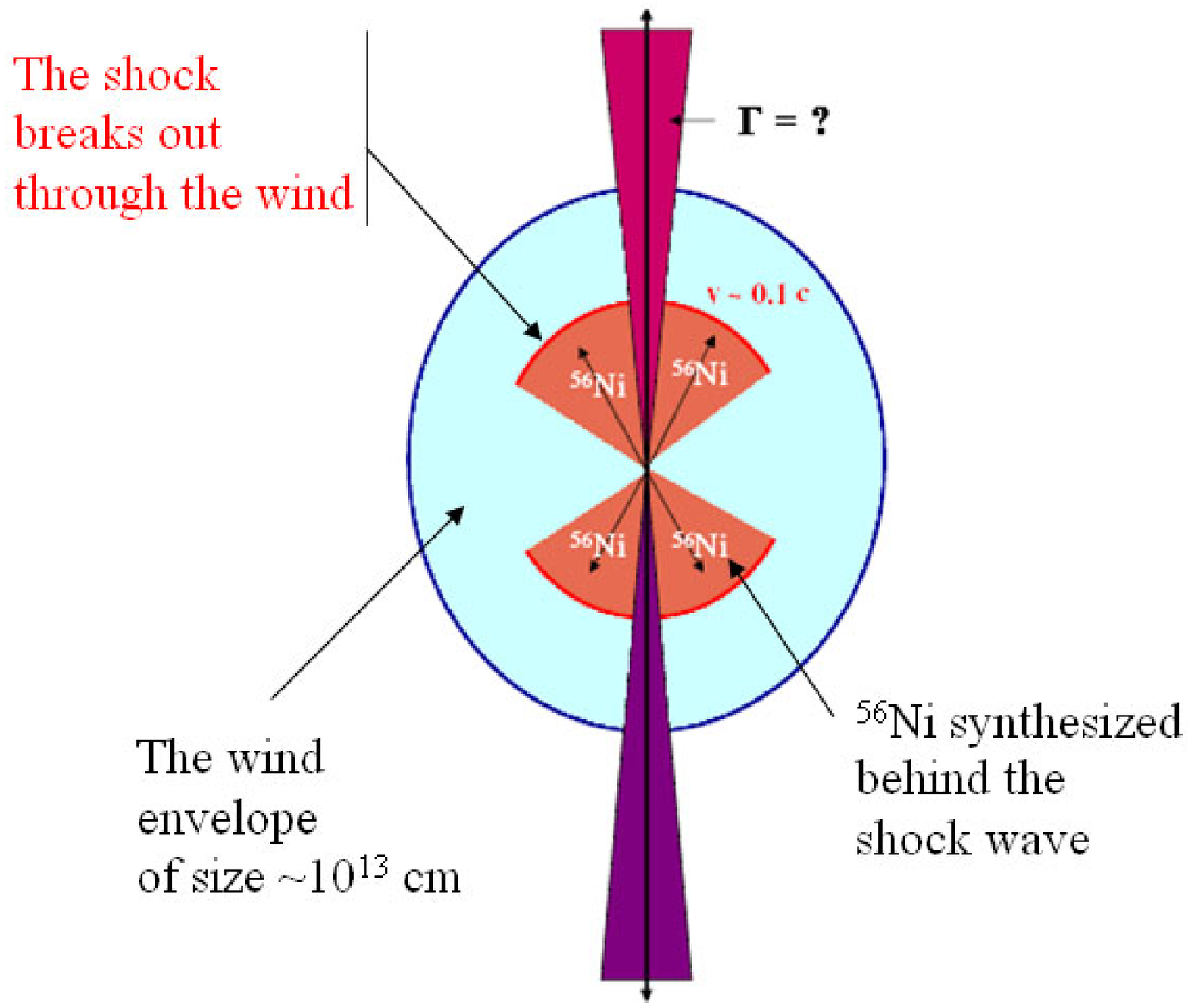}}
\caption{The schematic model of the asymmetric explosion of a GRB/SN progenitor.  
\label{f:Fig10}}
\end{figure}

More on puzzles of CCSNe and the GRB-SN connections see in Hjorth et al.(2011), though while it is not clear that the same mechanism that generates the GRB is also responsible for exploding the star. Or a strongly non-spherical explosion may be a generic feature of CCSNe of all types \citep{Leonard2006}. The searching for more Sp. confirmed pairs of GRBs (XRFs) and SNe in future observations is very important for understanding the nature of the GRB-SN connection, the nature of GRBs, and, eventually the mechanism itself of CCSNe explosion. In the context of aforesaid about GRB host galaxies, the accumulation of information on the GRB-SN connections can be considered as {\it the second result} of identification of GRBs. Now long-duration GRBs are identified with (may be) ordinary massive CCSNe. So, we have the {\it massive} star-forming in GRB hosts and {\it  massive} star explosions – CCSN/GRB. The question ''What kind of objects are the progenitors of GRBs?'' (see Figure \ref{f:Fig1}) becomes especially important at very high redshifts $z \gtrsim 10$. 

%----------- Section 4 -----------------------------

\section{The GRB rate and SFR at high redshifts and conclusions} 

Thus the long-duration GRBs are explosions associated to the collapse of short–lived massive stars ($\sim 30 M_{\odot}$), with peak emission at sub--MeV energies, where dust extinction is not an issue. Then the fluxes of these events can be detected eventually {\it from any redshift}. The death rate of massive short–lived stars (CCSNe rate) resembles their formation rate (massive SFR).  If the massive SFR straight proportional to the GRB formation rate (SFR $\sim$ GRBR), then the GRB rate (GRBR) can be used as a potential tracer of the massive SFR in the distant universe \citep{Ramirez2000}. If the GRBR $\sim$ SFR up to the high redshifts the questions arises: Is a fast decrease of SRF at $z > 4$ observed indeed, which is to be observed in cosmological models? Is there any difference between the GRBR and SFR beyond $z \sim 4$?  Below are some comments on SFR in galaxies with large redshifts (see Section 2) and on GRBs which serve now as powerful probes of the SFR at the highest redshifts.

        Indeed, as long-duration GRBs are associated with massive stars, therefore with regions of star formation, they (GRBs) are good candidates to study the SFR density. The GRB 090423 at $z = 8.2$ (see also Cucchiara et al.(2011)) on a photometric redshift of $z \sim 9.4$ for GRB 090429B) have further extended the redshift interval where the estimates of SFR evolution can be done, in a regime never explored before. Kistler et al.(2009) have compared SFR for different field galaxy samples with SFR derived from GRBs. It is based on the idea - the GRBR in galaxies is proportional to the SFR and that the ratio does not change with z. The normalization of GRB SFR density is done by taking the SFR density value at low redshift for which the density of the GRBR is estimated \citep{Ramirez2000}.

%---- Figure 11 ---------------------
\begin{figure}
\centerline{\includegraphics[width=10cm]{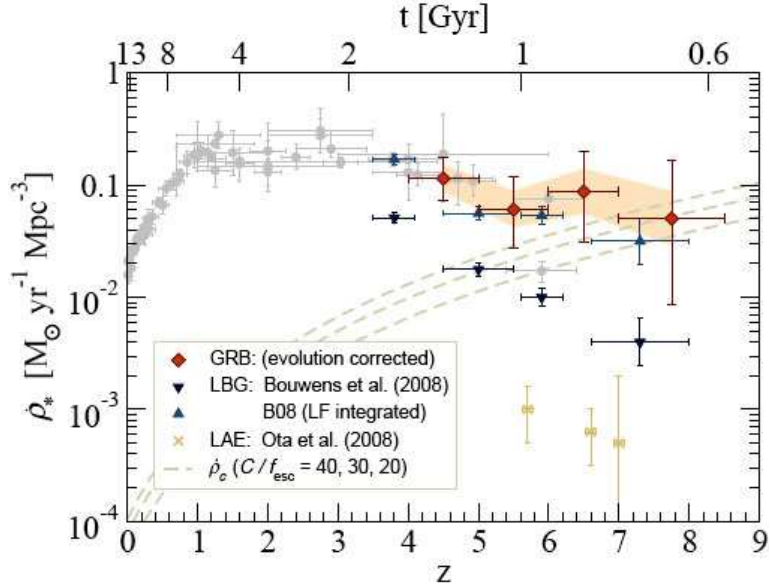}}
\caption{The cosmic star formation density from Kistler et al.(2009). Light circles are the data from Hopkins et al. (2006). Crosses are from Lyman-$\alpha$ emitters (LAE). Down and up triangles are Lyman-break galaxies (LBGs) for two UV luminosity functions.  The SFR inferred from GRBs (red diamonds) indicate {\it the strong contribution from small galaxies} generally not accounted for in the observed LBG luminosity function.
\label{f:Fig11}}
\end{figure}

       The SFR can remain high (see Figure 11) at least up to redshifts about 8 \citep{Kistler2009}. They also see no evidence for a strong peak in the SFR versus $z$. The agreement with direct observations, corrected for galaxies below detection thresholds, suggests that the GRB-based estimates incorporate the bulk of high-$z$ star formation down to the faint galaxies. At $z = 8$, GRB SFR density is consistent with LBG measurements after accounting for unseen galaxies at the faint-end UV luminosity function. This implies that not all star-forming galaxies at these $z$ are currently being accounted for in deep surveys. GRBs provide the contribution to the SFR from small galaxies -- {\it the typical GRB host at high redshift might be a small star forming galaxy}. Yan et al.(2009) come to analogous conclusions in their independent study of the high redshift SFR density with a glance at {\it the large number of low luminosity galaxies} (up to $M = 15.0$ mag.). May be no steep drop or fast decrease exists in the SFR  up to $z \sim 10$ indeed. In any case these results already affect the choice of cosmological parameters and star formation models on the cosmic star formation history in ΛCDM cosmological simulations \citep{Choi2009, Choi2011}.

And some conclusions:

1. The main conclusions resulting from the investigation of the GRB hosts (In point of fact, this is the first result of the GRB optical identification in 2001 with already known objects.): GRBs are identified with ordinary (or the most numerous in the universe at any $z$) galaxies up to $\sim 28$ st. magnitudes and more. The GRB hosts should not be special, but normal, faint, star-forming galaxies (the most abundant), detected at any $z$ just because a GRB event has occurred. The GRB hosts do not differ in anything from other galaxies with close redshifts: neither in colours, nor in spectra, the massive SRFs, and the metallicities. It means that these are generally star-forming galaxies (''ordinary'' for their redshifts) constituting the base of all deep surveys \citep{Bloom2001, Djorgovski2001, Frail2002, Sokolov2001a, Savaglio2006, Savaglio2008, Savaglio2009, Mannucci2011}. 

2. The second result of identification of GRBs: Now long-duration GRBs are identified with (may be) ordinary {\it massive} CCSNe. So, we have the {\it massive} star-forming in GRB hosts and {\it massive} star explosions --- CCSN/GRB. The search for differences between {\it nearby} SNe identified with GRBs and {\it distant} SNe which are to be identified with GRBs can be an additional observational cosmological test. We can ask a question analogous to that of 2001 on GRB hosts: Do GRB SNe differ from usual (local) SNe? Generally, what are redshifts at which CCSNe are quite different from local CCSNe? 

3. The main conclusions on an evolution for high redshifts (if GRBR $\sim$ SFR $\sim$ CCSN rate): The GRBs themselves and their hosts are already considered as a probe for studying processes of star-forming at cosmological distances up to $z \sim 10$ and more. As the universe is transparent to gamma rays up to $z \sim 10$ and more, in the process of study, {\it a new branch of observational cosmology} has arisen as a result of investigations of GRBs and their hosts.  Irrespective of specific models of the GRB phenomenon, it can be said now that when observing GRBs we observe SNe which, probably, are always related to the {\it relativistic collapse} of massive stellar cores in very distant galaxies. And at what $z (> 10-50?)$ are GRBs and massive CCSNe un-observable already? Now it could just be the main cosmological GRB-CCSN test.

%---- Figure table ---------------------
%\begin{figure}
%%
%\centerline{\includegraphics[width=11cm]{Fig_table.eps}}
%%
%\caption{The selected parameters of two host galaxies
%\label{f:Fig_table}}
%%
%\end{figure}

%---------- Bibliography -------------------

\label{lastpage}
%------------------------------------------------------------------------------%

\begin{thebibliography}{}

\bibitem[Abraham et al.(2004)Abraham et al.]{Abraham2004}
Abraham R.~G., Glazebrook K., McCarthy P.~J. et al. 2004, AJ, \textbf{127}, 2455

\bibitem[Blain et~al.(2000)Blain et al.]{Blain2000}
Blain A.~W. \&  Natarajan P., 2000, MNRAS, \textbf{312}, L39 (arXiv:astro-ph/9911468), see Figure 1.

\bibitem[Bloom et al.(2001)Bloom et al.]{Bloom2001}
Bloom J.S., Djorgovski S. G., Kulkarni S. R., 2001, ApJ, \textbf{554}, 678

\bibitem[Boella et al.(1997)Boella et al.]{Boella1997}
Boella G. et al., 1997, A\&AS, \textbf{122}, 299 

\bibitem[Branch et al.(2001)Branch et al.]{Branch2001}
Branch D.,  Baron E., Jeffery D.~J., 2001, as a Chapter in ''Supernovae and Gamma-Ray Bursters'' in Lecture Notes in Physics (Springer-Verlag), ed. K. W. Weiler., vol. 598, p.47-75 (arXiv:astro-ph/0111573)

\bibitem[Branch et al.(2002)Branch et al.]{Branch2002}
Branch, D. et al. 2002,  ApJ, \textbf{566}, 1005

\bibitem[Campisi et al.(2011)Campisi et al.]{Campisi2011}
Campisi M.~A. et al. 2011, arXiv:astro-ph/1105.1378v1, submitted MNRAS
 
\bibitem[Choi et al.(2009)Choi et al.]{Choi2009}
Choi J., Nagamine K., 2009, arXiv:astro-ph/0909.5425 (Effects of cosmological parameters and star formation models on the cosmic star formation history in LambdaCDM cosmological simulations.) 

\bibitem[Choi et al.(2011)Choi et al.]{Choi2011}
Choi J., Nagamine K., 2011, 4. arXiv:astro-ph/1101.5656 (On the inconsistency between the estimates of cosmic star formation rate and stellar mass density of high redshift galaxies.)

\bibitem[Costa et al.(1997)Costa et al.]{Costa1997} 
Costa E. et al.,  1997, Nature, \textbf{387}, 783

\bibitem[Cucchiara et al.(2011)Cucchiara et al.]{Cucchiara2011}
Cucchiara A., Levan A. J., Fox D. B et al., 2011, arXiv:astro-ph/1105.4915

\bibitem[Djorgovski et al.(2001)Djorgovski et al.]{Djorgovski2001}
Djorgovski S.~G., Kulkarni S.~R., Bloom J.~S., et al. 2001,  invited review in proc. ''Gamma-Ray Bursts in the Afterglow Era: 2nd Workshop'', eds. Costa E. et al., ESO Astrophysics Symposia, Berlin: Springer Verlag, p. 218 (arXiv:astro-ph/0107535)

\bibitem[Elmhamdi et al.(2006)Elmhamdi et al.]{Elmhamdi2006}
Elmhamdi A. et al., 2006, A\&A, \textbf{450}, 305 (arXiv:astro-ph/0512572)

\bibitem[Fernandez et al.(1999)Fernandez et al.]{Fernandez1999}
Fern\'andez-Soto A., Lanzetta, K.M., Yahil, A., 1999, ApJ., \textbf{513}, 34

\bibitem[Frail et al.(2002)Frail et al.]{Frail2002}
Frail, D. A. et al. 2002, ApJ, \textbf{565}, 829

\bibitem[Fynbo et al.(2008)Fynbo et al.]{Fynbo2008}
Fynbo L. P. U., et al. 2008, ApJ, \textbf{683}, 321

\bibitem[Fynbo et al.(2004)Fynbo et al.]{Fynbo2004}
Fynbo, J. U. P. et al. 2004, ApJ, \textbf{609}, 962

\bibitem[Galama et al.(1998)Galama et al.]{Galama1998}
Galama, T.J., Groot, P.J., van Paradijs, J.,q et al. 1998, ApJ., \textbf{497}, L13

\bibitem[Hjorth et al.(2011)Hjorth et al.]{Hjorth2011}
Hjorth J., Bloom J.~S, 2011, arXiv:astro-ph/1104.2274v1

\bibitem[Hogg et al.(1999)Hogg et al.]{Hogg1999}
Hogg, D.~W., Fruchter, A.~S., 1999, ApJ, \textbf{520}, 54

\bibitem[Hopkins et al.(2006)Hopkins et al.]{Hopkins2006}
Hopkins A.~M., Beacom J.~F., 2006, ApJ, \textbf{651}, 142

\bibitem[Kistler et al.(2009)Kistler et al.]{Kistler2009}
Kistler, M.~D. et al, 2009, ApJ, \textbf{705}, L104

\bibitem[Kurt et al.(2005)Kurt et al.]{Kurt2005}
Kurt V.~G. et al., 2005, Nuovo Cim., \textbf{C28}, 521 (arXiv:astro-ph/0505535)

\bibitem[Labb\'e~ et~ al.(2009)Labb\'e~ et~ al.]{Labbe2009}
Labb\'e I., Gonz\'alez V., Bouwens R.~J., et al., 2009, astro-ph/arXiv:0911.1356v5

\bibitem[Leonard et al.(2006)Leonard et al.]{Leonard2006}
Leonard D.~C., Filippenko A.~V. et al., 2006, arXiv:astro-ph/0603297

\bibitem[Levesque et al.(2010)Levesque et al.]{Levesque2010}
Levesque E~ M. et al., 2010, AJ, \textbf{140}, 1557

\bibitem[Mannucci et al.(2010a)Mannucci et al.]{Mannucci2010a}
Mannucci F. et al., 2010, MNRAS, \textbf{408}, 2115

\bibitem[Mannucci et al.(2010b)Mannucci et al.]{Mannucci2010b}
Mannucci F, Salvaterra R., Campisi M.~A., 2010, arXiv:astro-ph/1011.4506,  MNRAS, in press

\bibitem[Mannucci et al.(2011)Mannucci et al.]{Mannucci2011}
Mannucci F., Salvaterra R., Campisi M. A., 2011, arXiv:astro-ph/1011.4506v2, MNRAS, in press

\bibitem[Moskvitin et al.(2010)Moskvitin et al.]{Moskvitin2010}
Moskvitin et al., 2010, Astrophys. Bull., \textbf{65}, 132 (arXiv:astro-ph 1004.2633) 

\bibitem[Ramirez et al.(2000)Ramirez et al.]{Ramirez2000}
Ramirez-Ruiz E, Fenimore E.~E. \& Trentham N., 2000, arXiv:astro-ph/0010588, talk given at the CAPP2000 Conference on Cosmology and Particle Physics, Verbier, Switzerland, eds. J. Garcia-Bellido, R. Durrer and M. Shaposhnikov, (AIP,2001)

\bibitem[Salvaterra et al.(2009)Salvaterra et al.]{Salvaterra2009}
Salvaterra R. et al., 2009, Nature, \textbf{461}, 1258 

\bibitem[Savaglio et al.(2006)Savaglio et al.]{Savaglio2006}
Savaglio S., Glazebrook K., Le Borgne D., 2006, arXiv:astro-ph/0601528v2

\bibitem[Savaglio et al.(2008)Savaglio et al.]{Savaglio2008}
Savaglio S., Glazebrook K., Le Borgne D., 2008, arXiv:astro-ph/0803.2718v3

\bibitem[Savaglio et al.(2009)Savaglio et al.]{Savaglio2009}
Savaglio S., Glazebrook K., and Le Borgne D., 2009, ApJ, \textbf{691}, 182

\bibitem[Soderberg et al.(2005)Soderberg et al.]{Soderberg2005}
Soderberg, A. M. et al. 2005, ApJ, \textbf{627}, 877 (arXiv:astro-ph/0502553)

\bibitem[Sokolov et al.(1998)Sokolov et al.]{Sokolov1998}
Sokolov V.V.,  Kopylov A.I., Zharikov S.V., et al., 1998, A\&A, \textbf{334}, 117

\bibitem[Sokolov et al.(1999)Sokolov et al.]{Sokolov1999}
Sokolov V.V., Zharikov S.V., Baryshev Yu.V. et al., 1999, A\&A, \textbf{344}, 43
   
\bibitem[Sokolov et al.(2001a)Sokolov et al.]{Sokolov2001a}  
Sokolov V.V., Fatkhullin T., Castro-Tirado A.J.,  et al., 2001a, A\&A, \textbf{372}, 438   

\bibitem[Sokolov et al.(2001b)Sokolov et al.]{Sokolov2001b}
Sokolov V. V. et al., 2001b, Bull. Spec. Astrophys. Obs., \textbf{51}, 48-50

\bibitem[Sokolov et al.(2001c)Sokolov et al.]{Sokolov2001c}
Sokolov V., 2001c, in Proc. ''Gamma-Ray Bursts in the Afterglow Era: 2nd Workshop'', eds. Costa E. et al., ESO Astrophysics Symposia, Berlin: Springer Verlag, p. 136 (arXiv:astro-ph/0102492)

\bibitem[Sokolov et al.(2003)Sokolov et al.]{Sokolov2003}
Sokolov V. V. et al., 2003, Bull. Spec. Astrophys. Obs. \textbf{56}, 5-14  (arXiv:astro-ph/0312359)

\bibitem[Sonbas et al.(2008)Sonbas et al.]{Sonbas2008}
Sonbas et al., 2008, Astrophys. Bull., \textbf{63}, 228 (arXiv:astro-ph 0805.2657)

\bibitem[Tanvir et al.(2009)Tanvir et al.]{Tanvir2009} 
Tanvir N. et al., 2009, Nature, \textbf{461}, 1254 

\bibitem[van Paradijs et al.(1997)van Paradijs et al.]{Paradijs1997} 
van Paradijs J. et al., 1997, Nature, \textbf{386 }, 686

\bibitem[Woosley et al.(2006)Woosley et al.]{Woosley2006}
Woosley S., Heger A., 2006, AIP Conf.Proc., \textbf{836}, 398-407 (arXiv:astro-ph/0604131)

\bibitem[Yan et al.(2009)Yan et al.]{Yan2009}
Yan H. et al., 2009, arXiv:astro-ph/0910.0077v1, v2 and v3

\bibitem[Zafar et al.(2011)Zafar et al.]{Zafar2011}
Zafar T. et al., 2011, arXiv: astro-ph/1102.1469v2,  Fig.A.2 
    
\bibitem[Zeh et al.(2004)Zeh et al.]{Zeh2004}
Zeh A., Klose S., Hartmann D.~H., 2004, arXiv:astro-ph/0503311, in Proc of the 22nd Texas Symposium on Relativistic Astrophysics at Stanford. Stanford California, Dec. 13-17, 2004. ed. Chen P, et al.

\bibitem[Zharikov et al.(1998)Zharikov et al.]{Zharikov1998}   
Zharikov S~V., Sokolov V.~V, and Baryshev Yu.~V., 1998, A\&A, \textbf{337}, 356 

\bibitem[Zharikov et al.(1999)Zharikov et al.]{Zharikov1999}
Zharikov S.~V. and Sokolov V.~V., 1999, A\&ASS, \textbf{138}, 485

\end{thebibliography}
\end{document}